 \documentclass [12pt,a4paper      ]{article}
\usepackage{times}

\DeclareFontFamily{OT1}{times}{}
\DeclareFontShape {OT1}{times}{m }{n }{ <-> ptmr }{}
\DeclareFontShape {OT1}{times}{bx}{n }{ <-> ptmb }{}
\DeclareFontShape {OT1}{times}{m }{it}{ <-> ptmri}{}
\DeclareFontShape {OT1}{times}{bx}{it}{ <-> ptmbi}{}

\usepackage{amsmath}
\usepackage{amsfonts}
\usepackage{amssymb}
\usepackage{latexsym}

\usepackage{isridef}  

\renewcommand{\rmi}{i} 
\renewcommand{\rmd}{d} 

\begin{document}

\title{\bf\vspace{-2.5cm} Derivation of the potential, field,  and
                          locally-conserved charge-current density
                          of an arbitrarily moving point-charge }

\author{
         {\bf Andre Gsponer}\\
         {\it Independent Scientific Research Institute}\\ 
         {\it Oxford, OX4 4YS, England}
       }

\date{ISRI-06-04.20 ~~ December 31, 2008}

\maketitle

\begin{abstract}

The complete charge-current density and field strength of an arbitrarily accelerated relativistic point-charge are explicitly calculated.  The current density includes, apart from the well-established three-dimensional delta-function which is sufficient for its global conservation, additional delta-contributions depending on the second and third proper-time derivatives of the position, which are necessary for its local conservation as required by the internal consistency of classical electrodynamics which implies that local charge-conservation is an \emph{identity}.  Similarly, the field strength includes an additional delta-contribution which is necessary for obtaining this result.  The Li\'enard-Wiechert field and charge-current density must therefore be interpreted as nonlinear generalized functions, i.e., not just as distributions, even though only linear operations are necessary to verify charge-current conservation.  The four-potential from which this field and the conserved charge-current density derive is found to be unique in the sense that it is the only one reducing to an invariant scalar function in the instantaneous rest frame of the point-charge that leads to a point-like locally-conserved charge-current density.

~\\

\noindent 03.50.De ~ Classical electromagnetism, Maxwell equations

\end{abstract}

\section{Introduction}
\label{int:0} \setcounter{equation}{0}

Local conservation of the charge-current density four-vector $J_\MU$ is a necessary consequence of the antisymmetry of the electromagnetic field strength tensor $F_{\mu\nu}$.  Indeed, the identity
\begin{align}\label{int:1}
    \partial^{\mu} J_\MU \equiv 0,
\end{align}
derives from taking the four-divergence of both sides of Maxwell's inhomogeneous equation
\begin{align}\label{int:2}
    \partial^\nu F_{\mu\nu} =  - 4\pi J_\MU,
\end{align}
where the left-hand side vanishes after contraction because $F_{\mu\nu}$ is antisymmetric.  The vanishing of the four-divergence of the current-density is thus not an ordinary conservation law (i.e., a `weak' conservation law subject to the field equations being satisfied) but an \emph{identity} (i.e., a so-called strong conservation law) which has to be satisfied even if $J_\MU$ is a distribution rather than a smooth function.  This conclusion is absolutely general and should therefore be true for an arbitrarily moving relativistic point-charge, that is for the charge-current density of the Li\'enard-Wiechert field, which turns out not to be the case for the customary formulation of this field.

    In this paper we show that if the Li\'enard-Wiechert current is properly calculated, which implies that the Li\'enard-Wiechert field strength must be supplemented by an additional $\delta$-function-like field, local charge conservation is restored.  This conclusion is obtained by using only well known physical concepts, and a few basic results of distribution theory, but at the expense of some lengthy calculations whose details could not be given in the letter: \emph{The locally-conserved current of the Li\'enard-Wiechert field}, reference~\cite{GSPON2006C}.

    For ease of calculation we use the biquaternion formalism.  While this choice is a matter of convenience for the present paper, it is a necessity for subsequent publications because the spinor decomposition of four-vectors enables complicated four-dimensional integrations to be made very efficiently and in full generality.

   The general structure of this paper is the same as the letter \cite{GSPON2006C}, in which the tensor formalism was used throughout. Thus we begin in section~\ref{def:0} by setting forth our notations and definitions, including a number of basic quaternion definitions to make our paper self-contained.  A number of four-dimensional geometrical and kinematical identities needed in this and subsequent papers are derived, or else sufficient information is given to explain how they were derived.

   In section~\ref{cus:0} we recall the customary formulation of the Li\'enard-Wiechert potential and field, and show that the corresponding charge-current density is not conserved.  This requires calculations that are repeated in more details in the following sections, where it is also explained why the standard formulation does not lead to an identically-conserved current.

   The locally conserved charge-current density is derived in section~\ref{loc:0}. The method used consists of postulating a very general form for the four-potential, expressed in the causal coordinate system defined in section~\ref{def:0}, and to seek under which conditions it leads to a conserved point-like charge-current density.  This current density, i.e.,
\begin{align}\label{int:3}
    J_\MU=  \frac{e}{4\pi}
             \Bigl(   \frac{\dot{Z}_\MU}{\xi^2} 
                    + \frac{\ddot{Z}_\MU + 2\kappa K_\MU}{\xi}
                    - (2\kappa^2 + \chi) K_\MU
              \Bigr)  \delta(\xi)\Bigr|_{\tau=\tau_r},   
\end{align}
has a rather complicated structure: Kinematically, it depends on the three invariants $\xi, \kappa$, and $\chi$, as well as on the two four-vectors $\dot{Z}_\MU$ and $\ddot{Z}_\MU$, and on the biacceleration $\dddot{Z}_\MU$ through the invariant $\chi$; geometrically, on the angular variables through the null four-vector $K_\MU(\theta,\phi)$; and, distributionally, on Dirac's $\delta$-function and its first two derivatives.

   In section~\ref{str:0} a straightforward derivation of the conserved charge-current density is given.  This mathematically rigorous derivation is based on a full exploitation of Schwartz's structure theorem of distribution theory, which explains how $\delta$-functions arise as partial derivatives of continuous functions \cite{COURA1962-,CHOQU1982-}.  It will be seen that all is needed is a correct characterization of the nature of the singularity associated with a spin-less classical electron (namely a point-particle whose four-potential reduces to a scalar function in its instantaneous rest frame), and everything follows from a consistent application of Maxwell's theory.   This characterization consists of replacing the Li\'enard-Wiechert potential by the expression \cite{GSPON2004D,GSPON2008B,GSPON2006B}
\begin{align}\label{int:4}
    A_\MU = e \frac{\dot{Z}_\MU}{\xi}
              \Upsilon(\xi) \Bigr|_{\tau=\tau_r}, 
\end{align}
where the generalized function $\Upsilon(\xi)$ explicitly specifies how to consistently differentiate at $\xi=0$.  Similar forms of the potential of a classical point-charge have been discovered independently by Frank R.\ Tangherlini \cite{TANGH1962-}, as well as by others \cite{GSPON2004D}, in various contexts.  It then turns out that by starting from \eqref{int:4} to calculate the field $F_{\mu\nu}$ and current $J_\MU$ according to their standard definitions one immediately finds the locally conserved current \eqref{int:3}, provided all $\delta$-terms generated by differentiation are retained until the end of the calculation --- that is  provided $A_\MU, F_{\mu\nu}$ and $J_\MU$ are interpreted not just as distributions but as nonlinear generalized functions \cite{GSPON2008B,GSPON2006B}.

   In the conclusion, section~\ref{dis:0}, we discuss the consistency of our main results, equations \eqref{int:3} and \eqref{int:4}, with the standard formulations of Coulomb's potential, Lagrange's function, and Green's function of classical electrodynamics.  Finally, we summarize these results in the form of a theorem, stating that the potential \eqref{int:4} is unique in the sense that it is the only four-potential reducing to an invariant scalar function in the instantaneous rest frame that leads to a point-like locally-conserved charge-current density.

\section{Notations and definitions}
\label{def:0} \setcounter{equation}{0}

In this and related papers Hamilton's biquaternion (i.e., complexified quaternion) formalism is used instead of vector or tensor calculus.  This is because biquaternions provide concise and fully-general explicit expressions for all covariant quantities appearing in electrodynamics, including \emph{spinor} and \emph{null} quantities for which there are no explicit formula in ordinary vector or tensor calculus, e.g., equations \eqref{def:11} and \eqref{def:12} below.  This enables all calculations to be made explicitly and exactly (that is to obtain final results in closed form without using any approximation or limiting process) and within a fully-general and consistent four-dimensional framework that is eminently suited to the description of non-inertial motion in special-relativity.  

The same is true with regards to Cartan's differential forms, a formalism that is particularly well suited to abstract manipulations such as proving theorems, but not to explicit calculations.  For instance, with respect to equation \eqref{int:1}, the general form of the divergence is $\nabla\cdot J=\star d\star J$ where $\star$ is the Hodge dual --- whereas, from a strict mathematical point of view, $\nabla\cdot J=\partial^\mu J_\mu$ only for a Cartesian basis (even though it is commonly understood by physicists that this expression has to be properly adapted when working in a curvilinear coordinate system as in the present paper).  In response to this the biquaternion formalism provides, for the divergence in the four-dimensional space of special relativity, the general form used in equation \eqref{def:3} below, i.e.,  $\CON{\nabla} \scal J$, which in contradistinction to Cartan's expression $\star d\star J$ is explicit and directly amenable to calculations.  Indeed, in Cartesian coordinates Hamilton's operator $\nabla$ is given by equation \eqref{def:2} below, and in the curvilinear coordinate system used in the present paper by equation \eqref{def:27}.

A primer of quaternion methods in physics is given in reference~\cite{GSPON1993B}. However, as a quick start, we recall that the non-commutative algebra of quaternions consists of four-dimensional objects  $Q = s + \vec{v}$, were $s$ and $\vec{v}$ are the scalar and respectively vector parts of $Q$.  The quaternion product is defined as
\begin{align}\label{def:1}
    (a + \vec{a})(b + \vec{b}) \DEF ab - \vec{a}\cdot\vec{b} + a\vec{b} + b\vec{a} + \vec{a}\times\vec{b},
\end{align}
but detailed calculations requiring this formula are rarely needed since the crucial advantage of quaternions is that they can be manipulated as \emph{whole symbols}, therefore dispensing of complicated calculations at the component level, as in tensor and vector calculus.  This is particularly useful in relativity and electrodynamics where the Minkowski metric is obtained by allowing quaternion components to be complex numbers. (Hamilton introduced the prefix `bi' to denoted complexified objects, e.g., biquaternion, bivector, biscalar for complex number, etc.)

   For example, the four-gradient operator (i.e., the four-dimensional generalization of Hamilton's `nabla') and the four-vector current-density in equation \eqref{int:1} correspond to the biquaternions 
\begin{align}\label{def:2}
   \nabla = \frac{\partial}{\partial \rmi t}+\vec{\nabla},
        \qquad  \mbox{and} \qquad
    J = \rho -\rmi \vec{j}.
\end{align}
(We use Gaussian units and put the velocity of light in the vacuum equal to one.)  Equation \eqref{int:1} then becomes
\begin{align}\label{def:3}
    \CON{\nabla} \scal J \equiv 0,
\end{align}
where the unary operator $\CON{\A}$ denotes quaternion conjugation, namely the operation $\CON{(s + \vec v)} = s - \vec v$ such that $\CON{QR}=\CON{R} \,\, \CON{Q}$, and the binary operator $\scal$ as in $Q \scal R$ means taking the scalar part of the product $QR$.  As is shown in \cite{GSPON1993B}, products such as $Q\CON{R}S\CON{T}...$, where four-vectors and their quaternion conjugate alternate, are automatically covariant.  For example, $\CON{\nabla} \scal J$ is an invariant scalar.  Moreover, for any four-vector $Q$, the product $\CON{Q}Q = Q\CON{Q}$ is an invariant scalar: The square of its Minkowski norm.

   The electromagnetic field strength corresponds to the bivector $\vec{F} = \vec{E} + \rmi\vec{B}$, and Maxwell's inhomogeneous equation \eqref{int:2} translates to
\begin{align}\label{def:4}
    \nabla \vec{F} =  - 4\pi J.
\end{align}
By operating with $\CON{\nabla}$ on both sides, this equation leads to
\begin{align}\label{def:5}
    \CON{\nabla}\nabla \vec{F} =  - 4\pi \CON{\nabla}J,
\end{align}
where $\CON{\nabla}\nabla$ is a scalar operator: The d'Alembertian.  Therefore, since $\vec{F}$ is a vector, and because operating with a scalar operator on a vector can only lead to a vector, taking the scalar part on both sides of \eqref{def:5} automatically leads to \eqref{def:3}, the biquaternion form of the tensor identity \eqref{int:1}.  Consequently, in biquaternion formalism the necessity of the local conservation of $J$ is even more transparent than in tensor formalism:  It is immediately seen that this conservation is a consequence of the \emph{formal structure} of electrodynamics, and for this reason more fundamental than, for example, gauge invariance  \cite[p.~676]{JACKS2001-}.

   Finally, to end this brief introduction to quaternion methods, we give the definition of $\vec{F}$ in terms of the four-potential $A= \varphi - \rmi\vec{A}$, i.e.,
\begin{align}\label{def:6}
   \vec{F} \DEF \CON{\nabla} \vect A,
\end{align}
where the the binary operator $\vect$ as in $Q \vect R$ means taking the vector part of the product $QR$.

\subsection{Paul Weiss's causal coordinate system}

   The most important covariant null-quantity of field theory is, in tensor notation, the four-vector interval $R_{\mu}=X_{\mu}-Z_{\mu}$ between the position $Z_{\mu}$ of moving particle and the point $X_{\mu}$ at which the fields associated with that particle are observed.  Indeed, the nullity of the Minkowskian distance
\begin{align}\label{def:7}
    R_{\mu}R^{\mu} = |\vec x - \vec z|^2
                   - (t_x-t_z)^2  =0,
\end{align}
means that the points $X$ and $Z$ are positioned in such a way that a field observed at $X$ can be causally related to its source at $Z$, and vice-versa that a particle at $Z$ can be causally influenced by the field generated by another particle at $X$.  In the first case causality is implied by the condition that the time $t_z$ is preceding $t_x$, i.e.,
\begin{align}\label{def:8}
    t_z < t_x.
\end{align}
The corresponding root $t_r \DEF t_z$ of equation \eqref{def:7} is customarily called the `retarded time,' although `causal time' would be a better word to suggest that it is the time $t_r$ at which the source located at $\vec z$ produced the effect that is later observed at the location $\vec x$ at the time $t_x$.

  In biquaternion notation the observer's four-vector is $X=\rmi t_x + \vec x$ and that of the source $Z=\rmi t_z + \vec z$.  Equation \eqref{def:7} is then written
\begin{align}\label{def:9} 
           R \CON{R} = (X - Z) (\CON{X} - \CON{Z}) = 0.
\end{align}
If the world-line of the source is expressed in parametric form as a function $Z(\tau)$ of the proper time $\tau=\gamma t_z$, where  $\gamma = 1/\sqrt{1-\beta^2}$ is the  Lorentz factor and  $\vec \beta = \vec{v}/c$ the three-dimensional velocity, its four-velocity $\dot{Z}$ is given by
\begin{align}\label{def:10}
   \dot{Z}(\tau) = \gamma(1-\rmi\vec\beta) ,
     \qquad  \mbox{so that} \qquad
        \dot{Z}\CON{\dot{Z}} = 1,
\end{align}
where the dot corresponds to differentiation with respect to imaginary proper time, i.e., $\dot{Z} = {d}Z/{d\rmi\tau}$.

   Since biquaternions form an algebra, the four-velocity has a unique decomposition as a product 
\begin{align}\label{def:11}
   \dot{Z}(\tau) = \mathcal{B}  \mathcal{B}^+ ,
     \qquad  \mbox{such that} \qquad
        \mathcal{B}\CON{\mathcal{B}} = 1.
\end{align}
Here the operator $\A^+$ denotes quaternion biconjugation, namely the operation $(s + \vec v)^+ = \CON{(s + \vec v)}^* = s^* - \vec v^*$ such that $(QR)^+ = R^+ Q^+$, which combines quaternion and complex conjugations.  Eq.~\eqref{def:11} is the spinor decomposition of the four-velocity, and the unit biquaternion $\mathcal{B}(\tau)$ biunivocally defines (up to a factor $\pm 1$) the general Lorentz-boost to the frame in which the arbitrarily moving particle at the point $Z(\tau)$ is instantaneously at rest.

   In biquaternions the causal relationship \eqref{def:9} between the points $X$ and  $Z$ can be expressed by means of the spinor decomposition of the interval $X-Z$, discovered in 1941 by Paul Weiss \cite{WEISS1941-}, i.e.,
\begin{align}\label{def:12}
      X - Z = R =  \xi \mathcal{B} (\rmi + \vec{\nu}) \mathcal{B}^+.
\end{align}
Here $\xi$ is an invariant scalar and $\vec{\nu} = \vec{\nu}(\theta,\phi)$ a unit vector so that $\rmi + \vec{\nu}$, and consequently $R$, are null biquaternions. Equation~\eqref{def:12} is therefore an explicit parametrization of $R$ in terms of four variables: The invariants $\xi$ and $\tau$, and the two angles $\theta$ and $\phi$ characterizing the unit vector $\vec{\nu}$ in the instantaneous rest frame of the particle.  Thus, in that frame, i.e., when $\mathcal{B}=1$, the interval $X-Z$ reduces to the quaternion $\xi (\rmi + \vec{\nu})$, which shows that in that frame the vector $\xi \vec{\nu}$ corresponds to the ordinary radius vector $\vec x - \vec z$, and that the distance $\xi$ appears to an observer at rest in that frame as the ordinary distance $|\vec x - \vec z|$.  On the other hand, when $\mathcal{B}\neq 1$, Eq.~\eqref{def:12} provides a general parametrization of the null-interval $R$, that is, in geometrical language, of the `light-' or null-cone originating from $Z$.

   In the case where $Z$, $\xi$, $\vec{\nu}(\theta,\phi)$, and $\mathcal{B}$ are evaluated at the retarded proper time $\tau_r \DEF \gamma t_r$, Eq.~\eqref{def:12} defines a causal coordinate system erected at the retarded four-position $Z(\tau_r)$, i.e.,
\begin{align}\label{def:13}
    X(\tau,\xi,\theta,\phi)
      = Z + \xi \mathcal{B} (\rmi + \vec{\nu}) \mathcal{B}^+ \Bigr|_{\tau=\tau_r}.
\end{align}
Paul Weiss called it `retarded coordinate system,' but we prefer the term causal coordinate system.\footnote{Equations \eqref{def:12} and \eqref{def:13} describe the motion of a generally accelerated particle from the standpoint of an inertial frame that is momentarily at rest with respect to the particle.  In such a case the actual comoving frame of reference of the accelerating particle is noninertial, which implies that the object of anholonomity has to be introduced to properly define differentiation \cite{SCHOU1954-}.  In this paper this is taken care of by an intrinsic definition of the causal derivatives introduced in subsection (2.3) below.}  From the form of equations \eqref{def:12} and \eqref{def:13} it is immediately seen that $\xi$ can be expressed in terms of $R$ and $\dot{Z}$, namely as the scalar product
\begin{align}\label{def:14}
          \rmi\xi =   \dot{Z} \scal \CON{R} \Bigr|_{\tau=\tau_r},
\end{align}
which is the usual definition of the retarded distance $\xi$.  As $\xi$ is equal to the difference $\gamma(t_x - t_z) = \tau - \tau_r$, we also have $\tau_r= \tau - \xi$.

   In the present paper the spinor representation \eqref{def:12} of $R$ will not be used anymore.  Instead, to make all $\xi$ dependencies explicit, we introduce the `unit' null-quaternion $K$ such that
\begin{align}\label{def:15}
          K(\theta,\phi) \DEF R/\xi \Bigr|_{\tau=\tau_r},
\end{align}
where as above the condition  $\tau=\tau_r$ recalls that all quantities are evaluated at the retarded proper time $\tau_r$.  In the following, for simplicity,  this condition will be specified only for the main equations.

\subsection{Kinematical invariants}

In this paper we shall have to calculate expressions containing up to the fourth proper time derivative of $Z$.  This implies that apart from $\xi$ there will be three more invariants, namely the acceleration, biacceleration, and triacceleration invariants defined as:
\begin{align}
\label{def:16}
          \kappa &\DEF   \ddot{Z} \scal \CON{K},\\
\label{def:17}
      \rmi\chi   &\DEF \, \dddot{Z} \scal \CON{K},\\ 
\label{def:18}
          \zeta  &\DEF   \ddot{\ddot{Z}} \scal \CON{K}.
\end{align}
In the literature these invariants are sometimes defined as $\ddot{Z} \scal \CON{R}$,  $\dddot{Z} \scal \CON{R}$, etc.  As $R=\xi K$ this multiplies  our invariants by a factor $\xi$, which is better left explicit since $\xi$ is a coordinate (even though the combinations $\kappa\xi, \chi\xi$ and $\zeta\xi$ arise very often during calculations because of dimensionality).  

We shall also use a number of well-known kinematical identities that are easily found by iteratively derivating Eq.~\eqref{def:10}, and which yield the invariants:
\begin{align}
\label{def:19}
            \dot{Z} \scal & \CON{\dot{Z}} = \dot{Z}\CON{\dot{Z}} =  1,\\
\label{def:20}
           \ddot{Z} \scal & \CON{\dot{Z}} = 0,\\
\label{def:21}
          \dddot{Z} \scal & \CON{\dot{Z}} = -  \ddot{Z}\CON{\ddot{Z}}
                                          = -  \mathcal{A}^2,\\
\label{def:22}
         \ddot{\ddot{Z}} \scal & \CON{\dot{Z}} = -3 \dddot{Z}\scal\CON{\ddot{Z}},
\end{align}
where the invariant $\mathcal{A}$ is the magnitude of the acceleration.

\subsection{Causal derivatives}

When differentiating quantities depending on $X$ and $Z$ with respect to the coordinates of the point of observation $X$, one has to include the partial derivatives relative to both $X$ and $Z$ since they are related by the condition $R\CON{R}=0$, which insures causality.  Because $Z = Z(\tau_r)$ is an implicit function of $\tau$ this can be done with the help of a simple rule obtained by differentiating Eq.~\eqref{def:9}, i.e.,
\begin{align}\label{def:23} 
           2(dX - dZ) \scal \CON{R} =
           2(dX - d\rmi\tau_r \dot{Z}) \scal \CON{R} = 0.
\end{align}
Thus, 
\begin{align}\label{def:24} 
           d\rmi\tau_r = dX \scal \frac{\CON{R}}{\dot{Z} \scal \CON{R}}
           = -\rmi dX \scal \CON{K},
\end{align}
where we used \eqref{def:14} and the definition of $K$.  But, by definition of the total derivative,
\begin{align}\label{def:25} 
           d\rmi\tau_r =  dX \scal \CON{\nabla}\rmi\tau_r,
\end{align}
where $\nabla$ is the four-gradient operator \eqref{def:2}.  Therefore, comparing \eqref{def:24} and \eqref{def:25}, we find
\begin{align}\label{def:26} 
           \nabla \rmi\tau_r =  -\rmi K.
\end{align}
Consequently, for an expression $E = E(X,\tau)$, where the argument $X$ corresponds to an explicit dependence on $X$, and $\tau$ to the proper time, we have the covariant differentiation rule
\begin{align}\label{def:27}
   \nabla E(X,\tau) = \nabla E(X)
                    - \rmi K \dot{E}(\tau) \Bigr|_{\tau=\tau_r}. 
\end{align}

   We now use this rule to derive a number of simple relations which will be useful for calculating more complicated expressions in the following.  The first one is 
\begin{align}\label{def:28}
   \CON{\nabla} R = 4 + \rmi\CON{K}\dot{Z}. 
\end{align}
where we used $\CON{\nabla} X = 4$ which is obvious.  Further we have
\begin{align}\label{def:29}
         N \DEF \nabla \xi = (1-\kappa\xi) K - \rmi\dot{Z}, 
\end{align}
which defines the normal $N$ to the proper tube of radius $\xi$ surrounding the world-line, and which has the properties
\begin{align}
\label{def:30}
    \CON{N} \scal       {N} &=   1 - 2 \kappa\xi,\\
\label{def:31}
    \CON{N} \scal   \dot{Z} &=  -\rmi \kappa\xi,\\
\label{def:32}
    \CON{N} \scal  \ddot{Z} &=  \kappa(1-\kappa\xi),\\ 
\label{def:33}
    \CON{N} \scal \dddot{Z} &=  \rmi\mathcal{A}^2 + \rmi(1-\kappa\xi)\chi.
\end{align}

   Using the same method, i.e., the differentiation rule \eqref{def:27} and the previously obtained results, we also find
\begin{align}
\label{def:34}
      \CON{\nabla} K &=  - \CON{\nabla} \nabla \tau_r = \frac{2}{\xi},\\
\label{def:35}
      \CON{\nabla} N &= \CON{\nabla} \nabla \xi = (1-2\kappa\xi)\frac{2}{\xi}.
\end{align}
Similarly, we have the four-gradients
\begin{align}
\label{def:36}
    \nabla \kappa\xi &= \ddot{Z}  + \chi\xi K,\\
\label{def:37}
      \nabla \chi\xi &=  -\rmi\dddot{Z}  - (\mathcal{A}^2 + \zeta\xi) K.
\end{align}
The second of these involves a fourth proper-time derivative because $\zeta= \ddot{\ddot{Z}} \scal \CON{K}$. Accordingly, if we calculate the d'Alembertian of $\kappa\xi$, i.e., the gradient of equation \eqref{def:36}, we expect such a fourth derivative to arise.  However,
\begin{align}
\label{def:38}
         \CON{\nabla}\nabla \kappa\xi = 4 \chi. 
\end{align}
The reason for this is that when calculating the gradient of the second term in \eqref{def:36} we get
\begin{align}
\label{def:39}
    \CON{\nabla} \chi\xi K = 
                   \bigl(- \rmi \dddot{\CON{Z}}
                         - (\mathcal{A}^2 + \zeta) \CON{K} \bigl) K
                   + \chi\xi \CON{\nabla} K, 
\end{align}
so that the term containing $\zeta$ disappears because $\CON{K}K = 0$.  As will be seen, this is a general feature:  Causality insures that no derivatives of $Z$ beyond the third appear in any physically meaningful quantity considered in this paper.

\section{The customary formulation}
\label{cus:0} \setcounter{equation}{0}

The potential of the Li\'enard-Wiechert field can be obtained by a number of methods that are explained in many text books, e.g., \cite{SOMME1951-,LANDA1975-,JACKS1975-,BARUT1964-,INGAR1985-}. Most frequently it is obtained by working in the Lorenz gauge, and by means of a Green's function assuming that the point-charge can be represented by a three-dimensional $\delta$-function.  In biquaternion notation this source current-density can be written in the following covariant form   
\begin{align}\label{cus:1}
   J^{\rm S} = \frac{e}{4\pi} \dot{Z} \frac{1}{\xi^2} \delta(\xi),
\end{align}
whose normalization corresponds to the global (or integral) form of charge conservation, i.e.,
\begin{align}\label{cus:2}
  \int_{0}^{2\pi}  \rmd\phi
  \int_{0}^{\pi}   \rmd\theta \sin\theta 
  \int_{0}^{\infty}\rmd\xi~\xi^2 J =   e \dot{Z}.
\end{align}
The resulting potential has the remarkable simple form
\begin{align}\label{cus:3}
   A^{\rm LW} = e \frac{\dot{Z}}{\xi} \Bigr|_{\tau=\tau_r},
\end{align}
from which, one can calculate the electromagnetic field strength bivector $\vec{F} = \vec{E} + \rmi\vec{B}$ according to its definition, Eq.~\eqref{def:6}, which can be rewritten in equivalent form as
\begin{align}\label{cus:4}
   \vec{F} = \CON{\nabla} \vect A
           = \CON{\nabla} A  - \CON{\nabla} \scal A
           = B  - \psi,
\end{align}
where 
\begin{align}\label{cus:5}
              B \DEF \CON{\nabla} A,
              \qquad \mbox{and the scalar} \qquad
             \psi \DEF \CON{\nabla} \scal A,
\end{align}
is the invariant which is set equal to zero in the Lorenz gauge.

    Applying the rule \eqref{def:27} and using equation \eqref{def:29} it is easily found that
\begin{align}\label{cus:6}
        \CON{\nabla} A^{\rm LW} = - e \Bigl(
                              (1 - \kappa\xi)\frac{\CON{K}\dot{Z}}{\xi^2}
                           - \rmi\frac{(1 - \xi\CON{K}\ddot{Z})}{\xi^2}
                                         \Bigr)_{\tau=\tau_r},
\end{align}
of which the scalar part is, using equations \eqref{def:14} to \eqref{def:16} and being careful not to subtract potentially infinite quantities,
\begin{align}\label{cus:7}
        \psi^{\rm LW} = - e \Bigl(
                              (1 - \kappa\xi)\frac{\rmi}{\xi^2}
                  - \rmi\frac{(1 - \kappa\xi)}{\xi^2}
                                         \Bigr)_{\tau=\tau_r}
                = 0, \qquad \forall \xi \neq 0,
\end{align}
which shows that the Lorenz gauge, i.e., $\psi = 0$, is only satisfied when $\xi \neq 0$.  However, since the electromagnetic field strength is a vector, the fact that the scalar $\psi^{\rm LW}$ is undefined at $\xi = 0$ has no direct consequence on the field, even though the field $\vec{F}^{\rm LW}$ is derived from the same potential $A^{\rm LW}$, which just like $\vec{F}^{\rm LW}$ is undefined at $\xi = 0$.  Thus, taking the vector part of \eqref{cus:6} the Li\'enard-Wiechert field strength is
\begin{align}\label{cus:8}
        \vec{F}^{\rm LW}  = - e \Bigl(
                         (1 - \kappa\xi)\frac{\CON{K} \vect \dot{Z}}{\xi^2}
                                - \rmi  \frac{\CON{K} \vect \ddot{Z})}{\xi}
                       \Bigr)_{\tau=\tau_r}.
\end{align}

   Using again the rule \eqref{def:27} we can calculate the corresponding charge current density according to \eqref{def:4}.  However, since $\psi^{\rm LW} = 0$ for $\xi \neq 0$, it is easier to use \eqref{cus:4} and rewrite \eqref{def:4} as
\begin{align}\label{cus:9}
           J^{\rm LW} = - \frac{1}{4\pi} \nabla\CON{\nabla} A^{\rm LW},
                          \qquad \forall \xi \neq 0,
\end{align}
which, in fact, is the standard form used in the customary formulation assuming the Lorenz gauge.  Then using for $\CON{\nabla} A^{\rm LW}$ equation \eqref{cus:6}, we get after some calculations
\begin{align}\label{cus:10}
   J^{\rm LW} = \frac{e}{4\pi} \Bigl(
              2\rmi \bigl( \frac{1}{\xi^2} - \frac{1}{\xi^2} \bigr) \ddot{Z}
   - (1-2\kappa\xi) \bigl( \frac{2}{\xi^3} - \frac{2}{\xi^3} \bigr)  \dot{Z}
                 \Bigr).
\end{align}
In deriving this result we have been careful not to subtract potentially infinite quantities in order to see how $\delta$-functions, such as in the current density \eqref{cus:1}, arise.  Indeed, when $\xi \neq 0$ the current \eqref{cus:10} is zero, but since it is undefined for $\xi = 0$ one has to substitute an appropriate $\delta$-function for the ill-defined differences $1/\xi^n-1/\xi^n$.  The standard prescription for doing this is to use global charge conservation, i.e., equation \eqref{cus:2}, which implies that the expression $2/\xi^3-2/\xi^3$ should be replaced by $-\delta(\xi)/\xi^2$.  But there is no such constraint for the $1/\xi^2-1/\xi^2$ expression in the first term of \eqref{cus:10}.  However, if its origin is traced to the $1/\xi$ factor in the potential, one finds that it should be $+\delta(\xi)/\xi$ to be consistent with the $2/\xi^3-2/\xi^3$ expression.  With this prescription we get
\begin{align}\label{cus:11}
   J^{\rm LW} = \frac{e}{4\pi} \Bigl[\frac{\dot{Z}}{\xi^2}
                     +2\rmi \frac{\ddot{Z} - \kappa\dot{Z}}{\xi}
                 \Bigr] \delta(\xi) \Bigr|_{\tau=\tau_r}.
\end{align}
This current density differs from \eqref{cus:1} by the presence of two additional terms which depends on the acceleration.  However, when integrated over the whole three-space as in \eqref{cus:2}, $J^{\rm LW}$ yields the same total current $e\dot{Z}$ as $J^{S}$, because after multiplication by the volume element $\rmd\phi \, \rmd\theta \sin\theta\, \rmd\xi\, \xi^2$ the acceleration dependent terms does not contribute to the radial integral since $\xi\delta(\xi)=0$.  Unfortunately, if one tries to verify that the charge-current density $J^{\rm LW}$ is locally conserved, and consistently uses again the differentiation rule \eqref{def:27}, one finds
\begin{align}\label{cus:12}
   \CON{\nabla} \scal J^{\rm LW} = - \frac{e}{4\pi}
                        2\rmi \kappa \frac{1}{\xi^2}\delta(\xi) \neq 0.
\end{align}
This result is non-zero, and distributionally well-defined in the sense that the measure $\int \rmd\xi\,\xi^2 \partial^\MU J^{\rm LW}_\MU$ is finite.  The current density $J^{\rm LW}$ is therefore not locally conserved, because according to \eqref{int:1} the divergence \eqref{cus:12} should be {\it identically} zero.

  Moreover, if instead of $J^{\rm LW}$ one tries to verify the local conservation of the customary source current-density $J^{S}$ defined by \eqref{cus:1}, one also finds that it is not locally conserved.  Indeed, as this current is very simple, this is immediately seen by calculating its divergence, which consists of three terms, i.e.,  using equations \eqref{def:27},  \eqref{def:29}, \eqref{def:16}, and \eqref{def:31}
\begin{align}
\label{cus:13}
  \frac{4\pi}{e} \CON{\nabla} \scal J^{\rm S}
 &= - \rmi \CON{K} \scal \Bigl( \ddot{Z} \Bigr)\frac{1}{\xi^2} \delta(\xi)
\nonumber\\
   &~~~~+ \CON{N} \scal \dot{Z} \Bigl(-\frac{2}{\xi^3}\Bigr) \delta(\xi)
\nonumber\\
   &~~~~+ \CON{N}\scal\dot{Z} \frac{1}{\xi^2} \Bigl(\frac{-1}{\xi} \delta(\xi)
                                             \Bigr)\nonumber\\
   &= \Bigl( -\rmi\kappa  + 2 \rmi\kappa  + \rmi\kappa \Bigr)
     \frac{1}{\xi^2} \delta(\xi)
     \neq 0.   
\end{align}
Only in the inertial limit, that is for non-accelerated motion, are the current densities $J^{\rm LW}$ and $J^{\rm S}$ locally conserved.  This means that something is wrong in the customary formulation of the electrodynamics of an arbitrarily moving point-charge, or else that something is inconsistent and needs to be clarified.  For instance, to avoid any misunderstanding, it should be recalled that there is large class of conditions under which a simple current density like \eqref{cus:1} is locally conserved.  These comprise the case where instead of being related by a causality condition such as \eqref{def:9}, the $X$ and $Z$ variables are independent, e.g., \cite[p.~72]{LANDA1975-} and \cite[p.~139]{BARUT1964-}.

\section{The locally conserved charge-current density}
\label{loc:0} \setcounter{equation}{0}

In this section we derive the locally conserved current of an arbitrarily moving point-charge by making no other assumption than that its potential reduces to a scalar function $\varphi(\xi)$ in its instantaneous rest frame.  Thus, instead of taking for the potential the Li\'enard-Wiechert form, i.e., equation \eqref{cus:3}, we will assume
the general form
\begin{align}\label{loc:1}
    A \DEF e \dot{Z}(\tau) \varphi(\xi) \Bigr|_{\tau=\tau_r},
\end{align}
and seek under which conditions it leads to a conserved point-like current density.

    This method is motivated by the observation that any current distribution of the general form
\begin{align}\label{loc:2}
    J(X)= \frac{e}{4\pi} \Bigl( \frac{\dot{Z}}{\xi^2}
                              + \frac{S}{\xi}
                              + T \Bigr)\delta(\xi),
\end{align}
where $S$ and $T$ are any smooth four-vector functions, will satisfy global charge conservation because of the distributional identity $\xi\delta(\xi)=0$.  We therefore suppose that the absence of local current conservation could be due to an incorrect handling of the singularity at $\xi=0$, which should in fact lead to a current having a more complicated form than equations \eqref{cus:1} or \eqref{cus:11}.  Thus, to proceed step by step and be fully general, we  replace the $1/\xi$ factor in the potential \eqref{cus:3} by a function $\varphi(\xi)$ in \eqref{loc:1} that is finite and indefinitely differentiable except possibly at $\xi=0$.  This will lead us to a differential equation for a conserved current, of which we can study the solutions in terms of both regular functions and distributions.

   Moreover, in order to make gauge invariance explicit, we write Maxwell's equation for the current density $J$ directly in terms of $A$, i.e., from \eqref{def:4} and \eqref{cus:4},
\begin{align}\label{loc:3}
       \nabla \CON{\nabla} A 
     - \nabla \psi = - 4\pi J,
\end{align}
where $\psi$ is the invariant already defined in \eqref{cus:5}.

\subsection{General field-strength and gauge invariance}

    To find the field strength \eqref{cus:4} we first calculate $B$, which using \eqref{def:29} and \eqref{def:27}, is
\begin{align}\label{loc:4}
              B =       \CON{\nabla} A
                =       \CON{\nabla} e \dot{Z} \varphi
                = e     \CON{N}        \dot{Z} \varphi'
                - e \rmi\CON{K}       \ddot{Z} \varphi.
\end{align}
Using \eqref{def:31} and \eqref{def:16}, its scalar part is
\begin{align}\label{loc:5}
       \psi = -e \rmi \kappa \xi \varphi_1,
\end{align}
where
\begin{align}\label{loc:6}
       \varphi_1 \DEF \varphi' + \frac{1}{\xi}\varphi.
\end{align}
   The field strength, equation \eqref{cus:4}, is then obtained by subtracting the scalar $\psi$ from the biquaternion $B = \CON{\nabla} A$, what is equivalent to taking its vector part, i.e., 
\begin{align}\label{loc:7}
       \vec{F}  =      \CON{\nabla} \vect A
                = e    \CON{N} \vect  \dot{Z} \varphi'
                - e\rmi\CON{K} \vect \ddot{Z} \varphi,
\end{align}
which is gauge invariant.  Indeed, by adding to $A$ the four-gradient of any scalar function $f(X,Z)$ of the spatial coordinates and time, i.e., by making the substitution $A \rightarrow A + \nabla f$, we see that $B \rightarrow B + \CON{\nabla} \nabla f$ while $\psi \rightarrow \psi + \CON{\nabla} \nabla f$, so that $\vec{F} \rightarrow \vec{F}$ and consequently neither $\vec{F}$ nor $J$ depend on the choice of the gauge function $f$.

  Moreover, as $\psi$ has been subtracted from $B = \CON{\nabla} A$, the field strength $\vec{F}$ does not depend on $\psi$ nor on $\varphi_1$, only on $\varphi$ and $\varphi'$ separately, as can be seen in \eqref{loc:7}.

\subsection{General charge-current density}

   To find the general current density \eqref{loc:3}, we begin by calculating $\nabla B$, what is simpler to do if it is rewritten as $e \nabla \CON{\nabla} \dot{Z} \varphi$.  Thus,
\begin{align}\label{loc:8}
    \nabla B = e\Bigl( \nabla \CON{\nabla} \dot{Z} \Bigr) \varphi 
             + e\Bigl( \nabla \varphi \Bigr) \Bigl( \CON{\nabla} \dot{Z} \Bigr)
             + e\nabla        \Bigl( \CON{\nabla} \varphi \Bigr)        \dot{Z}
             + e\Bigl( \nabla \CON{\nabla} \varphi \Bigr) \dot{Z},
\end{align}
where we have used the `mobility' of the scalar function $\varphi$ to put it at the right place for the differentiations which are made within the ranges indicated by the parentheses.  The four terms in this expression are
\begin{align}
\label{loc:9}
       \Bigl( \nabla \CON{\nabla} \dot{Z} \Bigr) \varphi
           = -  \rmi \Bigl( \nabla \CON{K} \dot{Z} \Bigr) \varphi 
           = -2 \rmi \frac{1}{\xi} \ddot{Z} \varphi,\\
\label{loc:10} 
       \Bigl( \nabla \varphi \Bigr) \Bigl( \CON{\nabla} \dot{Z} \Bigr)
           = ( N \varphi' )(-\rmi\CON{K}\ddot{Z})
           = - \dot{Z}\CON{K}\ddot{Z}\varphi',\\
\label{loc:11} 
           \nabla  \Bigl( \CON{\nabla} \varphi \Bigr) \dot{Z}
           = -\rmi K  (\CON{N}\varphi') \ddot{Z}
           = - K \dot{\CON{Z}}\ddot{Z}\varphi',\\
\label{loc:12} 
           \Bigl( \nabla \CON{\nabla} \varphi \Bigr) \dot{Z}
           = \Bigl( \varphi' \nabla \CON{N}  + N \CON{N} \varphi''\Bigr) \dot{Z}
           = (1 - 2 \kappa\xi) (\varphi'' + \frac{2}{\xi} \varphi') \dot{Z},
\end{align}
where in the course of the calculations the identities of section~\ref{def:0} were used, including the relation $\CON{K}K = 0$ which enabled several simplifications.  Combining these four terms, and noting that $ \dot{Z}\CON{K} + K\dot{\CON{Z}} = 2 \dot{Z} \scal \CON{K} = 2\rmi$, we find
\begin{align}\label{loc:13}
    \nabla B = - e 2\rmi  \ddot{Z} \varphi_1
               + e (1-2\kappa\xi) \dot{Z}  \varphi_2.
\end{align}
Here,
\begin{align}\label{loc:14}
     \varphi_2(\xi) = \varphi'' + \frac{2}{\xi} \varphi',
\end{align}
has been introduced as an expression distinct from $\varphi_1$, even though it could be expressed in terms of  $\varphi_1$ as  $\varphi_2 = \varphi_1' + \varphi_1/\xi$. This is because we want to avoid any subtraction of potentially undefined terms, and therefore keep everything expressed in function of $\varphi$ and its derivatives as they arise.  Moreover, as seen in \eqref{loc:12}, $\varphi_2$ arises in the expression of the d'Alembertian of the scalar function $\varphi(\xi)$, which corresponds to the field in spin-zero electrodynamics, and is therefore of interest on its own.

   We next calculate $\nabla \psi$, which according to \eqref{loc:5} is
\begin{align}\label{loc:15}
     \nabla \psi = 
     - e \rmi \varphi_1 \nabla \kappa \xi
     - e \rmi \kappa\xi N \varphi_1'.
\end{align}
Using \eqref{def:29} and \eqref{def:36} this can be rewritten as
\begin{align}\label{loc:16}
     \nabla \psi = 
     - e \rmi (\ddot{Z} + \chi \xi K) \varphi_1
     - e \rmi \kappa\xi\bigl( (1-\kappa\xi)K - \rmi \dot{Z} \bigr) \varphi_1',
\end{align}
which shows that whereas $\nabla B$ given by \eqref{loc:13} is function of $\dot{Z}$ and $\ddot{Z}$,  $\nabla \psi$, and thus the current density $J$ given by \eqref{loc:3}, are moreover function of $\dddot{Z}$ through the invariant $\chi$. 

   Finally, combining \eqref{loc:13} and \eqref{loc:16}, the general form of the charge-current density is
\begin{align} \label{loc:17}
          -4\pi J  =  & + e (1-2\kappa\xi) \dot{Z}     \varphi_2\nonumber\\
                      & - e \rmi (\ddot{Z} -\chi\xi K) \varphi_1\nonumber\\
                      & + e \rmi \bigl( (1-\kappa\xi)K - i \dot{Z}
                                 \bigr) \kappa\xi \varphi_1'. 
\end{align}
Comparing with \eqref{loc:12}, the first line is seen to come from the d'Alembertian of $\varphi(\xi)$, which is also at the origin of $\varphi_2(\xi)$.  On the other hand, the second and third lines, as well as $\varphi_1(\xi)$, come from both $\nabla B$ and $\nabla \psi$.

In the inertial limit, $\dot{Z} = \Cst$, this current-density reduces to the simple expression
\begin{align}\label{loc:18}
              -4\pi J  =  + e \dot{Z}  \varphi_2,
\end{align}
which when $\varphi=1/\xi$ is just the Green's function usual source current density \eqref{cus:1}.  On the other hand, the Li\'enard-Wiechert current-density \eqref{cus:11} is obtained by ignoring the contributions coming from $\psi$, that is by taking equation \eqref{loc:13} equal to $-4\pi J$ and assuming $\varphi=1/\xi$.

\subsection{Divergence of general charge-current density}

   Since we have the general form of the charge-current density we can now explicitly verify that $\CON{\nabla} \scal J \equiv 0$, i.e., equation \eqref{int:1}.  Of course, we expect that to be true if there is no  algebraic error in our calculations.  But it is nevertheless important to do it since we want to make sure that nothing special arises in terms of handling ill-defined quantities, and also because we want to know under which precise conditions the charge-current density is locally conserved.

  However, since the calculation of $\CON{\nabla} \scal J$ is tedious, and basically consists of repetitively using the previously used methods and identities, we will just give the result, which is
\begin{align}\label{loc:19}
      \CON{\nabla} \scal J  =
          & + 2 e \rmi\chi\xi 
            \Bigl( (\varphi_1'  + \frac{1}{\xi}\varphi_1)
                                             - \varphi_2 \Bigr)\nonumber\\ 
          & + e \rmi \kappa\xi (1 - 2\kappa\xi)
            \Bigl( (\varphi_1'' + \frac{2}{\xi}\varphi_1') 
                 - (\varphi_2'  + \frac{1}{\xi}\varphi_2 ) \Bigr). 
\end{align}

   As can be seen, this expression contains only the three invariants $\xi$, $\kappa$, and $\chi$.  There is therefore no dependence on the fourth proper-time derivative of $Z(\tau)$, despite that the initial expression, the potential \eqref{loc:1}, depends on the first derivative through $\dot{Z}$ and $\rmi\xi = \CON{R}\scal\dot{Z}$.  As seen at the end of section~\ref{def:0}, this is because causality insures that no derivatives of $Z$ beyond the third appear in any physically meaningful quantity considered in this paper.  

   On the other hand, concerning $\varphi(\xi)$, the condition of validity of equation \eqref{loc:19} is, as expected, that it is three times differentiable, provided $\xi \neq 0$.  Hence, if we assume such a function for $\varphi(\xi)$, and calculate the expressions in the two big parentheses, we find that they are both identically zero, implying that the charge-current density is indeed conserved for any $\xi$ except possibly for $\xi = 0$.  However, if $\varphi(\xi)$ is a distribution rather than a regular function, which means that the derivatives of $\varphi(\xi)$ will be defined even at $\xi = 0$, the current density $J$ will also be locally conserved, but the condition $\xi \neq 0$ will no more be required.

   Finally, in order to illustrate the importance of not having made any simplification before getting $\CON{\nabla} \scal J$ in its final form as in \eqref{loc:19}, we remark that the first line of this equation would not be there had we expressed $\varphi_2$ in terms of $\varphi_1$, a possibility that was mentioned below equation \eqref{loc:14}.

\subsection{Locally-conserved general charge-current density}

    We finally come to the Li\'enard-Wiechert case, and therefore specialize to $\varphi=1/\xi$. 

   For $\xi\neq 0$ we have then
\begin{align}\label{loc:20} 
  \varphi_1(\xi) &= -\frac{1}{\xi^2} + \frac{1}{\xi^2} = 0,
  \quad \mbox{and} \quad
  \varphi_2(\xi) &= +\frac{2}{\xi^3} - \frac{2}{\xi^3} = 0,
\end{align}
which imply that all terms in \eqref{loc:17} are zero, including $\xi\varphi_1'= \xi\varphi_2-\varphi_1$.  The current density $J$ is then everywhere zero, except at $\xi =0$ where it is undefined. We therefore interpret $\varphi_1(\xi)$ as a distribution, and use the theorem stating that \emph{a distribution which has its support only in one point, say the origin, is a linear combination of the $\delta$-function and its derivatives up to a certain order} \cite[p.~784]{COURA1962-}, \cite[p.~443]{CHOQU1982-}. Thus
\begin{align}\label{loc:21} 
  \forall \xi \geqslant 0,  \qquad \varphi_1(\xi)
      = \frac{1}{\xi} \delta(\xi),
\end{align}
which because of dimensionality comprises a single $\delta$-function, and whose  normalization will turn out to be consistent with \eqref{cus:2}.  Similarly, we also interpret $\varphi_2(\xi)$ as a distribution, and, since by \eqref{loc:14} it can be expressed in terms of $\varphi_1(\xi)$, we get\footnote{Incidentally, comparing with the discussion above equation \eqref{cus:11}, we have rigorously justified the prescription that led to the non-locally-conserved charge-current density \eqref{cus:11}.}
\begin{align}\label{loc:22} 
  \forall \xi \geqslant 0,  \qquad \varphi_2(\xi)
      = \varphi_1' + \frac{1}{\xi} \varphi_1
      = -\frac{1}{\xi^2} \delta(\xi).
\end{align}
It remains to substitute \eqref{loc:21} and \eqref{loc:22} into \eqref{loc:17}, and the locally-conserved charge-current density \eqref{loc:3} is finally found to be
\begin{align}\label{loc:23}
    J=  \frac{e}{4\pi}
             \Bigl( \frac{\dot{Z}}{\xi^2} 
                    + \rmi \frac{\ddot{Z} + 2\kappa K}{\xi}
                    - \rmi (2\kappa^2 + \chi) K
              \Bigr)  \delta(\xi)\Bigr|_{\tau=\tau_r}.
\end{align}

   This result leads to several observations:
\begin{enumerate}

\item The current density $J$ is much more complicated than the simple current \eqref{cus:1}:  It depends directly on the three invariants $\xi, \kappa$, and $\chi$, as well as on the two four-vectors $\dot{Z}$ and $\ddot{Z}$; indirectly on the biacceleration $\dddot{Z}$ through the invariant $\chi$; and, finally, on the angular variables through the null four-vector $K(\theta,\phi)$.

\item The dependence on the third derivative of $Z$ is consistent with the Lorentz-Dirac equation and with the Schott expression of the self-force, in which $\dddot{Z}$ also appears, because the self-interaction force involves a product of $J$ with the self-field.

\item Equation \eqref{loc:23} has the most general distributional form of \eqref{loc:2}, in accord with the theorem cited above equation \eqref{loc:21}.

\item With $\varphi_1$ given by \eqref{loc:21} the invariant $\psi$ defined by \eqref{loc:5} is
\begin{align}\label{loc:24} 
         \psi = -e \rmi \kappa \delta(\xi).
\end{align}
Thus, the gauge can be the Lorenz gauge $\psi=0$ only for $\xi\neq 0$ when $\kappa\neq 0$, i.e., when the acceleration is non-zero.  

\item The equation $\varphi_1(\xi) = \varphi' + \varphi/\xi = 0, \forall \xi \neq 0,$ has only one solution: $1/\xi$. This singles out the corresponding potential as being the only one such that the current density of a point-charge is conserved and thus given by \eqref{loc:23}.

\end{enumerate}

\section{Straightforward derivation}
\label{str:0} \setcounter{equation}{0}

While the previous section's derivation of the locally-conserved current of an arbitrarily moving point-charge is rigorous, it has one important shortcoming: The derivation basically consisted of using distribution theory to find the correct form of that current, without having explicitly specified the forms of the potential and field leading to it,  even though we know from section~\ref{cus:0} that something is deficient in their customary formulation since they do not lead to the locally conserved current.  Indeed, what we expect from a consistent application of Maxwell's theory is that once we have properly formulated the potential or the field, the current should derive from either of them in a straightforward manner, that is without having to `mend' some ill-defined result to get the correct one.

   From a mathematical perspective, this shortcoming consists of having considered just $\varphi_1$ and $\varphi_2$ as distributions, whereas according to equations \eqref{loc:6} and \eqref{loc:14} they both derive from $\varphi$ which appears in the potential \eqref{loc:1} already, and which therefore should also be defined in the context of distribution theory.  In other words, we have not looked at the origin of the $\delta$-functions arising in $\varphi_1$ and $\varphi_2$, which according to Schwartz's structure theorem of distribution theory must arise from the partial differentiation of some continuous function.  In fact, the generating function leading to these $\delta$-functions is easily found.  This is because, in three-dimensional notation, the retarded distance Eq.~\eqref{def:14} reads
\begin{align}\label{str:1}
              \xi = |\vec{x} - \vec{z} | \gamma(1-\vec{\rho}\cdot\vec{\beta}),
\end{align}
where $\vec{\rho}$ is the unit vector in the direction of $\vec{x} - \vec{z}$. The retarded distance is therefore proportional to an absolute value, and for that reason has a discontinuous derivative when $\vec{x} \rightarrow \vec{z}$, i.e., at $\xi=0$.  Thus, when calculating derivatives of functions expressed in the coordinate system $\{ \xi,\theta, \phi \}$, one must carefully distinguish between the coordinate $\xi$ and the absolute value $|\xi|$, a crucial observation that was first made by Tangherlini \cite[p.~511--513]{TANGH1962-}.

   Consequently, as is explained in details in references \cite{GSPON2004D,GSPON2008B,GSPON2006B}, the potential of an arbitrarily moving accelerated point-charge must be written
\begin{align}\label{str:2}
    A = e \frac{\dot{Z}}{\xi} \Upsilon(\xi) \Bigr|_{\tau=\tau_r},
\end{align}
where $\Upsilon(\xi)$ is the generalized function defined as\footnote{Intuitively, $\Upsilon(r)$ can be seen as equivalent to the sign function ${\rm sgn}(r)$ for $r \geq 0$.}
\begin{align} \label{str:3}
   \Upsilon(r) \DEF 
         \begin{cases}
         \text{undefined}   &   r < 0,\\
                      0     &   r = 0,\\
                     +1     &   r > 0,
         \end{cases}
   \qquad\text{and}\qquad
     \frac{d}{dr}\Upsilon(r) = \delta(r).
\end{align}

   When the definition \eqref{def:6} and the causal differentiation rule \eqref{def:27} are now used to calculate the field strength starting from the potential \eqref{str:2}, the corresponding current density \eqref{def:4} is directly found to be the conserved one, i.e., equation \eqref{loc:23}.  However, instead of the customary Li\'enard-Wiechert field, equation \eqref{cus:8}, the field strength is now
\begin{align}
\label{str:4}
         \vec{F} = &- e \Bigl(
                          (1 - \kappa\xi)\frac{\CON{K} \vect \dot{Z}}{\xi^2}
                                  - \rmi \frac{\CON{K} \vect \ddot{Z})}{\xi}
                      \Bigr)\Upsilon(\xi) \Bigr|_{\tau=\tau_r},\\
\label{str:5}
                   &+ e \Bigl(
                           (1 - \kappa\xi)\frac{\CON{K} \vect \dot{Z}}{\xi}
                      \Bigr)\delta(\xi) \Bigr|_{\tau=\tau_r},
\end{align}
which apart from the presence of the $\Upsilon$-function multiplying $\vec{F}^{\rm LW}$ on the first line, has an additional $\delta$-like contribution on the second.  Indeed, starting from the general form of the field $\vec{F}$ given by \eqref{loc:7}, and taking for $\varphi$ the function implied by the potential \eqref{str:2}, it suffice to substitute
\begin{align}\label{str:6}
    \varphi = \frac{1}{\xi}\Upsilon(\xi),
            \qquad  \mbox{and} \qquad
    \varphi' = -\frac{1}{\xi^2}\Upsilon(\xi) + \frac{1}{\xi}\delta(\xi),
\end{align}
in \eqref{loc:7} to obtain (\ref{str:4}--\ref{str:5}) after use of the identity $\CON{N} \vect \dot{Z} = (1-\kappa\xi) \CON{K} \vect \dot{Z}$.

   Since both the $\Upsilon$-factor and the $\delta$-like contribution in $\vec{F}$ are necessary to obtain the current density satisfying the local conservation identity \eqref{int:1}, it is clear that the customary $\vec{F}^{\rm LW}$ cannot lead to such a current.  In fact, it is by calculating the current density immediately from the potential as in \eqref{loc:3} --- that is by ignoring that the field strength could be different from the customary one --- that after many unsuccessful attempts the author discovered the conserved current density in July 2003.

   Indeed, the $\delta$-like field on the second line of (\ref{str:4}--\ref{str:5}) is absolutely necessary since it carries an essential part of the information about the nature of the singularity at $\xi=0$, which is not simply due to the divergence of $1/\xi$ when $\xi \rightarrow 0$,  as in the customary formulation, but to that of  $1/|\xi|$ where the absolute value leads to the expression $\Upsilon(\xi)/\xi$ in the potential \eqref{str:2}.

  As this $\Upsilon$-function in the potential \eqref{str:2} is at the origin of the $\delta$-functions arising in the conserved current, we can come back on the functions $\varphi_1$ and  $\varphi_2$ defined in the previous section, and observe that starting from $\varphi=\Upsilon/\xi$ we have $\varphi_1 = \delta/\xi$ and $\varphi_2 = \delta'/\xi$.  Comparing to equations \eqref{loc:21} and \eqref{loc:22}, we also have $\varphi_1 = -\delta_{\rm D}'$ and $\varphi_2 = -\delta_{\rm D}''$, where $\delta_{\rm D}$ is the usual `Dirac $\delta$-function' which has the property $\delta_{\rm D}' = -\delta_{\rm D}/\xi$ characteristic of Schwartz's distribution-theory.  However, since $\varphi_1$ and $\varphi_2$, and thus $\varphi$, enter the conservation equation \eqref{loc:19} --- which is an identity --- that property is not necessary for insuring local charge conservation.  This means that any generalized function such that $\Upsilon' = \delta_{\rm C}$ (where $\delta_{\rm C} \neq \delta_{\rm D}$ is for example a Colombeau generalized function \cite{GSPON2006B}) is sufficient to have local charge conservation provided $\delta_{\rm C}'$ and $\delta_{\rm C}''$ exist.

  The major implication of the necessity of the $\delta$-contribution \eqref{str:5} to get an identically conserved charge-current density is that the field (\ref{str:4}--\ref{str:5}) cannot be interpreted as a distribution in which the $\delta$-contribution \eqref{str:5} can be ignored if the field is twice differentiated to calculate the divergence \eqref{int:1}.  The reason is that every successive differentiation introduces new terms which must be retained in order to get a result that is identically zero.\footnote{This can be illustrated by the following example: the function $f(\xi)=\xi^{-1}\delta(\xi)$ yields zero when evaluated on any test function $T(\xi)$ in $\mathbb{R}^3$.  On the other hand, $f'(\xi)=-2\xi^{-2}\delta(\xi)$ yields $-2T(0)\neq 0$ so that differentiation and evaluation on test functions do not commute in general.}  The field (\ref{str:4}--\ref{str:5}) and the current density \eqref{loc:23} must therefore be interpreted as nonlinear generalized function, such as Colombeau functions \cite{GSPON2006B}, in which all terms can be or can become significative in further calculations, even though none of the nonlinear properties of such functions have to used in the present paper.

\section{Discussion}
\label{dis:0} \setcounter{equation}{0}

In this paper we have derived the proper formulation of the potential of an arbitrarily moving point-charge, equation~\eqref{str:2}, which leads to the current-density \eqref{loc:23} that is locally-conserved, and to the field strength \eqref{str:4} which contains an $\Upsilon$-function factor and a $\delta$-function term that are absent in the customary Li\'enard-Wiechert form of that field.  Since the reasoning leading to these results is fully general, and therefore valid in the inertial limit, the Coulomb potential in the rest frame of a point-charge should thus be written \cite{GSPON2004D,GSPON2008B,GSPON2006B}
\begin{align} \label{dis:1}
          \varphi_C(\vec r) \DEF e\frac{1}{r}\Upsilon(r),
\end{align}
which implies that the corresponding Coulomb field
\begin{align} \label{dis:2}
       \vec E_C(\vec r) = -\vec \nabla \varphi_C
                        =  e\frac{\vec r}{r^3}\Upsilon(r) 
                        -  e\frac{\vec r}{r^2}\delta(r), 
\end{align}
also has a $\delta$-function term that is absent in the standard formulation of that field.  As shown in \cite{GSPON2008B}, this $\delta$-term implies that the self-energy of a point charge is not a $1/r$ distribution but a $\delta^2(r)$ singularity, so that the self-energy of a point charge is fully concentrated at $r=0$.

    As the necessity to supplement the customary Li\'enard-Wiechert and Coulomb fields by a $\delta$-function term is quite unexpected, it is important to stress that this conclusion was reached while remaining entirely within the realm of standard Maxwell theory, that is without making any modification to that theory.  The only thing that was done was to carefully analyze the nature of the singularity at the position of the point-charge, and to use distribution theory to formulate the implications stemming from it.  Moreover, while the derivation of the conserved current was somewhat indirect in section~\ref{loc:0}, its straightforward derivation from the potential
\begin{align}\label{dis:3}
    A = e \frac{\dot{Z}}{\xi} \Upsilon(\xi) \Bigr|_{\tau=\tau_r}, 
\end{align}
in section~\ref{str:0}, and the rigorous mathematical justification of that potential, make that the replacement of the Coulomb potential and field by the expressions \eqref{dis:1} and \eqref{dis:2} is of absolute physical necessity.

    Another aspect of the present paper may look somewhat surprising: distribution theory has since long been applied to classical electrodynamics by a number of researchers, and no modification to the Li\'enard-Wiechert potential and field was found to be necessary, see, e.g., \cite{TAYLO1955-, ROWE-1978-}.  The reason is that these researchers did not question the standard form of the Li\'enard-Wiechert potential and of the fields deriving from it: they simply postulated that they were distributions in order to consistently work with their singularities at $\xi=0$.  On the other hand, what we have done in this paper was to uncover the proper form of the potential leading to a point-charge current-density that is locally conserved --- which implied that just like in nonlinear generalized function theory all terms obtained by differentiation had to be kept until the end of the calculation in order to get an identity.

  In doing so we have in fact discovered more than we had hoped for, namely a striking result that can be phrased in the form of a unicity theorem:

\begin{quote}
\noindent {\bf Theorem [Unicity of the four-potential of an arbitrarily moving point-charge]}
{\it
 ~~ The only potential of the form
\begin{align}\label{dis:4}
        A = e \varphi(\xi)\dot{Z}(\tau) \Bigr|_{\tau=\tau_r}, 
\end{align}
where $\varphi(\xi)$ is a scalar function, such that the charge-current-density
\begin{align}\label{dis:5}
        -4\pi J  = \nabla (\CON{\nabla} \vect A),
\end{align}
is a locally conserved $\delta$-like nonlinear generalized function, i.e., such that
\begin{align}
\label{dis:6}
      \CON{\nabla} \scal J  &\equiv 0, \qquad \forall \xi,\\
                     J(\xi) &= 0, \qquad \forall \xi \neq 0,
\end{align}
is the unique solution of the equation
\begin{align}\label{dis:7}
         \varphi'(\xi) + \frac{1}{\xi} \varphi(\xi)
                       = \frac{1}{\xi} \delta(\xi),
\end{align}
namely
\begin{align}\label{dis:8}
        \varphi(\xi) = \frac{1}{\xi} \Upsilon(\xi),
\end{align}
provided all terms obtained by differentiation are kept until the end of the calculation. 
}
\end{quote}

   Finally, since the formulation of the potential and field of an arbitrarily moving point-charge presented in the present paper is new, there is an apparent contradiction with the fact that the customary Li\'enard-Wiechert formulation is an agreement with so many applications of classical electrodynamics.

   There is however no contradiction, since, on the contrary, the results of this paper are in full agreement with the fundamental principles of electrodynamics and mechanics.   For instance, if the conserved current \eqref{loc:23} is introduced in a Lagrange function as the scalar product $\CON{J} \scal A_{\rm ext}$ with the potential of an external field $A_{\rm ext} \in \mathcal{C}^\infty(\mathbb{R}^4)$, the differences between that current and the current $J^{\rm S}$ defined by equation \eqref{cus:1} have in general no influence since they disappear upon integration over the whole space.  The same is true for the derivation (which also involves an integration) of the potential of an arbitrarily moving point charge, i.e., equations \eqref{cus:3} or \eqref{str:2}, by means of a Green's function. 

   Thus, the principles of physics imply that the position $Z$ and velocity $\dot{Z}$ of a point-charge are sufficient to determine the potential of its field, while the precise formulation of that potential given by \eqref{str:2} is necessary to determine the complete field and conserved current-density, which include terms that are function of $\ddot{Z}$ and $\dddot{Z}$.  In other words, while $J^{\rm S}$ is sufficient as a source to determine uniquely the potential of an arbitrarily moving point-charge, the conserved current $J$ deriving from this potential can be very different from $J^{\rm S}$.  Indeed, without the factor $e$, $J^{\rm S}$ is merely a velocity distribution associated to the world-line of the point-charge.

    In conclusion, the formulation presented in this letter will make little difference in most engineering-type applications of classical electrodynamics.  Indeed, as can be seen by studying a number of examples, the instances in which the full details of the current density \eqref{loc:23} are strictly necessary, and the additional $\delta$-contributions in the fields (\ref{str:4}--\ref{str:5}) and \eqref{dis:2} are essential, include fundamental problems like calculating the interaction of a point-charge with itself \cite{GSPON2008B}, and similar nonlinear problems in which classical electrodynamics is apparently not consistent.  For example, the self-force on an arbitrarily moving point-charge is given by $F_{\mu\nu}  J^\NU$ where $F_{\mu\nu}$ is given by (\ref{str:4}--\ref{str:5}) and $J^\NU$ is the full conserved current density \eqref{loc:23}, i.e., such that $ \partial^\nu F_{\mu\nu} =  - 4\pi J_\MU$, not the customary charge-current density \eqref{cus:1}:  a frequent confusion which is a cause of known difficulties.  The resolution of such internal contradictions is the subject of several forthcoming publications \cite{GSPON2007A,GSPON2007B}.

   Finally, we stress that the methods used in present paper to derive the potential, field, and charge-current density of a point-charge can be extended to the derivation of those of an electron, i.e., a point-charge endowed with a magnetic dipole moment $\vec \mu$ and spin \cite{GSPON2008B}, whose four-potential expressed in the spinor notation of section~\ref{def:0} is \cite{GSPON2004C}
\begin{align}\label{dis:9}
        A_{\rm electron} = \mathcal{B} \bigl(
                 \frac{e}{\xi} - i\frac{\vec{\nu} \times \vec{\mu} }{\xi^2}
                              \bigr) \mathcal{B}^+\Upsilon(\xi).
\end{align}

\section{Acknowledgements}
\label{ack:0}

Extended correspondence with  Profs.\ William R. Davis,  Jean-Pierre Hurni, and Frank R. Tangherlini is greatfully acknowledged.


\newpage

\section{References}
\label{biblio}

\end{document}